\documentclass[%
onecolumn,
preprint,
 amsmath,amssymb,
 aps,    prc
]{revtex4-2}

\usepackage{graphicx}% Include figure files
\usepackage{dcolumn}% Align table columns on decimal point
\usepackage{bm}% bold math

\usepackage{multirow}

\usepackage{rotating}

\begin{document}

\title{Quasi-bound state in the $\bar{K}NNN$ system}% Force line breaks with \\

\author{N.V. Shevchenko}%
 \email{shevchenko@ujf.cas.cz}
\affiliation{%
 Nuclear Physics Institute, 25068 \v{R}e\v{z}, Czech Republic
}%

\date{\today}% It is always \today, today,

\begin{abstract}
The paper is devoted to the $\bar{K}NNN$ system, which is an exotic system consisting
of an antikaon and three nucleons. Dynamically exact four-body Faddeev-type equations
were solved and characteristics of the quasi-bound state in the system were evaluated.
Three antikaon-nucleon and three nucleon-nucleon potentials were used, so 
the dependence of the four-body pole positions on the two-body interaction models was studied.
The resulting binding energies $B^{\rm Chiral}_{K^-ppn} \sim 30.5 - 34.5$ MeV
obtained with chirally motivated and $B^{\rm SIDD}_{K^-ppn} \sim 46.4 - 52.0$ MeV obtained with
phenomenological antikaon-nucleon potentials are close to those obtained for the $K^- pp$
system with the same $\bar{K}N$ and $NN$ potentials, while the four-body widths
$\Gamma_{K^- ppn} \sim 38.2 - 50.9$ MeV are smaller.
\end{abstract}

%\pacs{13.75.Jz, 36.10.Gv}
%13.75.Jz: kaon-baryon interactions
%36.10.Gv: Mesonic atoms and molecules, hyperonic atoms and molecules

\keywords{few-body physics, antikaon-nucleon systems, four-body Faddeev equations}

\maketitle

\section{Introduction}

Attractive nature of $\bar{K}N$ interaction lead to suggestions, that quasi-bound states
can exist in few-body systems consisting of antikaons and nucleons~\cite{AY1}.
In particular, a deep and relatively narrow quasi-bound state was predicted  in the
lightest three-body $\bar{K}NN$ system~\cite{AY2}. Many theoretical calculations of the system
were performed after that using different methods and inputs. 
All of them agree, that the quasi-bound state really exists in spin-zero state of $\bar{K}NN$,
usually denoted as $K^- pp$, but predict quite different binding energies and widths of the state.

The experimental situation is unsettled as well: several candidates for the $K^- pp$ state
were reported by different experiments \cite{KppExp1,KppExp2,KppExp3}, while other
experiments left the matter unsettled \cite{KppExp4,KppExp5}. However, the measured
binding energies and especially decay widths of the state differ from each other and are far from all
theoretical predictions. The most recent results by J-PARK E15 experiment
\cite{KppExpE15-1,KppExpE15-2} for the binding energy are comparable to some theoretical
predictions, but the width of the $K^- pp$ quasi-bound state is still too large.

In recent years we performed a series of calculations of different states of the
three-body $\bar{K}NN$ and $\bar{K}\bar{K}N$ systems \cite{review}, using dynamically exact
Faddeev-type equations in AGS form with coupled $\bar{K}NN$ and $\pi \Sigma N$ channels.
In particular, we evaluated $K^- pp$ quasi-bound state binding energy and width using three
different models of $\bar{K}N$ interaction. The same was done for the $\bar{K}\bar{K}N$ system.
We also demonstrated, that there is no quasi-bound states, caused by pure strong interactions,
in spin one state of $\bar{K}NN$ system, which is $K^- np$. In addition, we calculated
near-threshold amplitudes of $K^-$ elastic scattering on deuteron.  Finally, we evaluated $1s$
level shift in kaonic deuterium, which is an atomic state, caused by presence of the strong
$\bar{K}N$ interaction in comparison to the pure Coulomb state.

Four-body $\bar{K}NNN$ system is another system with strangeness, which could lead more light
on the question of antikaon-nucleon interaction. Some theoretical calculations of the quasi-bound
state were already performed \cite{AY1,BGL,OHHMH,ir}, but more accurate calculations are needed.
We calculated binding energy and width of the state using four-body Faddeev-type equations
by Grassberger and Sandhas \cite{4AGS}. Only these dynamically exact equations in momentum
re\-pre\-sen\-tation can treat energy-dependent $\bar{K}N$ potentials, necessary for the this system,
exactly. We used our two-body antikaon-nucleon potentials, constructed for the three-body AGS
calculations \cite{review}, as the input.

\section{Four-body Faddeev-type equations}

The three-body Faddeev-type equations in Alt-Grassberger-Sandhas form
\begin{equation}
\label{3AGS}
 U_{\alpha \beta}(z) = (1-\delta_{\alpha \beta}) G^{-1}_0(z) + 
 \sum_{\gamma=1}^3 (1-\delta_{\alpha \gamma}) T_{\gamma}(z) G_0(z) U_{\gamma \beta}(z)
\end{equation}
define the three-body transition operators $U_{\alpha \beta}(z)$, which describe process
$\beta + (\alpha \gamma) \to \alpha + (\beta \gamma)$. The $G_0(z)$ in Eq.~(\ref{3AGS})
is three-body Green function, Faddeev partition indices $\alpha, \beta = 1,2,3$
simultaneously define a particle ($\alpha$) and the remained pair ($\beta \gamma$),
$\alpha \ne \beta \ne \gamma$. The operator $T_{\alpha}(z)$ is a two-body $T$-matrix,
describing interaction in the ($\beta \gamma$) pair.

A separable potential $V_{\alpha}$ leading to a separable $T$-matrix
\begin{equation}
\label{VTsep}
 V_{\alpha} = \lambda_{\alpha} |g_{\alpha} \rangle \langle g_{\alpha}| \to 
 T_{\alpha}(z) = |g_{\alpha} \rangle \tau_{\alpha}(z) \langle g_{\alpha}|
\end{equation}
allows to write the three-body AGS equations in the form
\begin{equation}
\label{3AGSsep}
 X_{\alpha \beta}(z) = Z_{\alpha \beta}(z) + 
 \sum_{\gamma=1}^3 Z_{\alpha \gamma}(z) \tau_{\gamma}(z) X_{\gamma \beta}(z)
\end{equation}
with new transition $X_{\alpha \beta}$ and kernel $Z_{\alpha \beta}$ operators,
defined by
\begin{eqnarray}
 X_{\alpha \beta}(z) &=&
  \langle g_{\alpha} | G_0(z) U_{\alpha \beta}(z) G_0(z) | g_{\beta} \rangle, \\
 Z_{\alpha \beta}(z) &=& (1-\delta_{\alpha \beta})
   \langle g_{\alpha} | G_0(z) | g_{\beta} \rangle.
\end{eqnarray}
Here for simplicity the one-term separable potentials Eq.(\ref{VTsep}) used, while
in general $V_{\alpha}$ can consist of $N$ terms.

%------------------------------------------------------------------------------
The four-body Faddeev-type Grassberger-Sandhas equations were derived in Ref.\cite{4AGS}
\begin{eqnarray}
\nonumber
 U_{\alpha \beta}^{\sigma \rho}(z) &=&
  (1-\delta_{\sigma \rho}) \delta_{\alpha \beta} G_0^{-1}(z) T_{\alpha}^{-1}(z) G_0^{-1}(z) + \\
  &{}& \qquad \sum_{\tau,\gamma} (1-\delta_{\sigma \tau}) U_{\alpha \gamma}^{\tau}
   G_0(z) T_{\gamma}(z) G_0(z) \, U_{\gamma \beta}^{\tau \rho}
\label{4AGS}
\end{eqnarray}
In addition to the free Green function $G_0(z)$, which now acts in four-body space, and the
two-body $T$-matrix $T_{\alpha}(z)$, three-body $U_{\alpha \beta}^{\tau}(z)$ and
four-body $U_{\alpha \beta}^{\sigma \rho}(z)$ operators enter the system Eq.(\ref{4AGS}).
The high indices $\sigma,\rho,\tau$ define a partition, which could be $3+1$ or $2+2$ type,
while the low indices $\alpha, \beta$ define two-body subsystems
of the particular three-body subsystem, denoted by the high index.

If the separable potentials Eq.~(\ref{VTsep}), leading to the corresponding separable $T$-matrices,
is used,
the system Eq.~(\ref{4AGS}) can be rewritten in the same way, as the three-body one. The new
system of equations
\begin{equation}
\label{4AGSsepV}
 \bar{U}^{\sigma \rho}_{\alpha \beta}(z) = (1-\delta_{\sigma \rho})  
    ( \bar{G_0}^{-1} )_{\alpha \beta}(z) + 
 \sum_{\tau,\gamma,\delta} (1-\delta_{\sigma \tau}) \bar{T}^{\tau}_{\alpha \gamma}(z) 
   (\bar{G_0})_{\gamma \delta}(z) \bar{U}^{\tau \rho}_{\delta \beta}(z)
\end{equation}
contains new operators
\begin{eqnarray}
 &{}& \bar{U}^{\sigma \rho}_{\alpha \beta}(z) = 
  \langle g_{\alpha} | G_0(z) U^{\sigma \rho}_{\alpha \beta}(z) G_0(z) | g_{\beta} \rangle, \\
 &{}& \bar{T}^{\tau}_{\alpha \beta}(z) = 
  \langle g_{\alpha} | G_0(z) U^{\tau}_{\alpha \beta}(z) G_0(z) | g_{\beta} \rangle, \\
\label{G0bar}
 &{}& (\bar{G_0})_{\alpha \beta}(z) = \delta_{\alpha \beta} \tau_{\alpha}(z).
\end{eqnarray}

It is seen that the four-body system with separable potentials  Eq.~(\ref{4AGSsepV}) looks
similarly to Eq.~(\ref{3AGS}), which describes a three-body system with arbitrary potentials.
This analogy can be used for further modification of the equations. Namely, if the three-body
$T$-matrices $\bar{T}^{\tau}_{\alpha \beta}(z)$ in Eq.~(\ref{4AGSsepV}) are presented in
a separable form
\begin{equation}
 \bar{T}^{\tau}_{\alpha \beta}(z) =  | \bar{g}^{\tau}_{\alpha} \rangle 
  \bar{\tau}^{\tau}_{\alpha \beta}(z)   \langle \bar{g}^{\tau}_{\beta} |,
\end{equation}
the four-body equations  Eq.~(\ref{4AGSsepV})  can be rewritten as
\begin{equation}
\label{4AGSsepVT}
 \bar{X}^{\sigma \rho}_{\alpha \beta}(z) = \bar{Z}^{\sigma \rho}_{\alpha \beta}(z) + 
 \sum_{\tau,\gamma,\delta} \bar{Z}^{\sigma \tau}_{\alpha \gamma}(z) 
     \bar{\tau}^{\tau}_{\gamma \delta}(z) \bar{X}^{\tau \rho}_{\delta \beta}(z)
\end{equation}
with new transition $\bar{X}^{\sigma \rho}_{\alpha \beta}$ and kernel
$\bar{Z}^{\sigma \rho}_{\alpha \beta}$ operators,
defined by
\begin{eqnarray}
 \bar{X}^{\sigma \rho}_{\alpha \beta}(z) &=&
  \langle \bar{g}^{\sigma}_{\alpha} | (\bar{G_0})_{\alpha \alpha}(z) 
    \bar{U}^{\sigma \rho}_{\alpha \beta}(z) 
   (\bar{G_0})_{\beta \beta}(z) | \bar{g}^{\rho}_{\beta} \rangle, \\
\label{Zbar}
 \bar{Z}^{\sigma \rho}_{\alpha \beta}(z) &=& (1-\delta_{\sigma \rho})
   \langle \bar{g}^{\sigma}_{\alpha} | (\bar{G_0})_{\alpha \beta}(z) | \bar{g}^{\rho}_{\beta} \rangle.
\end{eqnarray}
We solved the four-body equations Eq.~(\ref{4AGSsepVT}) with separable two-body $T$-matrices
being an input and three-body $T$-matrices $\bar{T}_{\alpha \beta}^{\tau}$ represented in
separable form.

%%%%%%%%%%%%%%%%%%%%%%%%%%%%%%%%%%%%%%%%%%%%%%%%%%%
\section{Separable three-body amplitudes}
\label{separab.sec}

We used $\bar{K}N$ and $NN$ potentials, which are separable ones by construction. 
Therefore, separable versions of three-body and 2+2 amplitudes, entering the
equations~(\ref{4AGSsepVT}), should be constructed. These amplitudes are described by 
three-body AGS equations Eq.(\ref{3AGSsep}), which being written in momentum basis  for $s$-wave
interactions have a form
\begin{equation}
\label{3AGSppz}
 X_{\alpha \beta}(p,p';z) = Z_{\alpha \beta}(p,p';z) +
 \sum_{\gamma = 1}^3 4 \pi
 \int_{0}^{\infty} Z_{\alpha \gamma}(p,p'';z) \, \tau_{\gamma}(p'';z) \, X_{\gamma \beta}(p'',p';z) p''^2 dp''.
\end{equation}
Here $p,p'$ and $z$ are relative momenta and three-body energy.
It is possible to evaluate eigenvalues $\lambda_n$ and eigenfunctions $g_{n \alpha}(p;z)$ of the
system Eq.(\ref{3AGSppz}) from
\begin{equation}
\label{gHS}
 g_{n \alpha}(p;z) = \frac{1}{\lambda_n} \,
 \sum_{\gamma = 1}^3 4 \pi
 \int_{0}^{\infty} Z_{\alpha \gamma}(p,p';z) \, \tau_{\gamma}(p';z) \, g_{n \gamma}(p';z) p'^2 dp'
\end{equation}
with normalization condition
\begin{equation}
\label{normHS}
 \sum_{\gamma = 1}^3 4 \pi
 \int_{0}^{\infty} g_{n \gamma}(p';z) \, \tau_{\gamma}(p';z) \, g_{n' \gamma}(p';z) p'^2 dp' = -\delta_{nn'}.
\end{equation}
Knowledge of the eigenvalues and eigenfunction allows us to write down Hilbert-Schmidt expansion
(HSE) of the kernel functions $Z_{\alpha \beta}$:
\begin{equation}
 Z_{\alpha \beta}^{\rm HSE}(p,p';z) = - \sum_{n=1}^{\infty} \lambda_n g_{n \alpha}(p;z) g_{n \beta}(p';z),
\end{equation}
which leads to the separable three-body amplitude
\begin{equation}
 X_{\alpha \beta}^{\rm HSE}(p,p';z) = - \sum_{n=1}^{\infty} \frac{\lambda_n}{1-\lambda_n} 
        g_{n \alpha}(p;z) g_{n \beta}(p';z).
\end{equation}
Since the kernel function $Z_{\alpha \beta}$ and three-body amplitude $X_{\alpha \beta}$
are energy-dependent functions, being off energy shell in the four-body equations Eq.(\ref{4AGSsepVT}),
it is necessary to solve eigenequations Eq.(\ref{gHS}) with normalization condition Eq.(\ref{normHS})
for every value of the three-body energy during the four-body calculations. It is timely
consuming work. Instead of this we used Energy Dependent Pole Expansion/Approximation
(EDPE/EDPA) method, suggested in~\cite{EDPE} specially for the four-body GS equations.
It needs solution of the eigenequations~Eq.(\ref{gHS}) only once, for a fixed energy $z_{\rm fix}$. Usually it is chosen
to be the binding energy $z_{\rm fix}=E_B$ if a bound state in the system exists or $z_{\rm fix}=0$ if not. 
After that energy dependent form-factors
\begin{equation}
 \label{gEDPE}
g_{n \alpha}(p;z) = 
 \sum_{\gamma = 1}^3 4 \pi
 \int_{0}^{\infty} Z_{\alpha \gamma}(p,p';z) \, \tau_{\gamma}(p';z_{\rm fix}) \, g_{n \gamma}(p';z_{\rm fix}) p'^2 dp'
\end{equation}
and propagators
\begin{eqnarray}
\label{thetaEDPE}
 \left( \Theta(z) \right)^{-1}_{mn} \, = \,
 \sum_{\gamma = 1}^3 4 \pi
 \int_{0}^{\infty} g_{m \gamma}(p';z) \, \tau_{\gamma}(p';z_{\rm fix}) \, g_{n \gamma}(p';z_{\rm fix}) p'^2 dp'  \\
\nonumber
-  \sum_{\gamma = 1}^3 4 \pi
 \int_{0}^{\infty} g_{m \gamma}(p';z) \, \tau_{\gamma}(p';z) \, g_{n \gamma}(p';z) p'^2 dp'
\end{eqnarray}
are calculated. 
Finally, the separable three-body amplitude can be written in a form
\begin{equation}
\label{X_EDPE}
 X_{\alpha \beta}^{\rm EDPE}(p,p';z) = \sum_{m,n=1}^{\infty}  g_{m \alpha}(p;z) \, \Theta_{mn}(z) \, g_{n \beta}(p';z).
\end{equation}
If only one term is taken in the sums in~Eq.(\ref{X_EDPE}), the Energy Dependent Pole Expansion
turns into Energy Dependent Pole Approximation. It is seen, that EDPE method needs only one solution
of the eigenvalue equations Eq.(\ref{gHS}) and calculations of the integrals Eqs.(\ref{gEDPE},\ref{thetaEDPE})
after that. According to the authors, the method converges faster than Hilbert-Schmidt expansion, it
is accurate already with one term.

Three-body form-factors  $g_{\alpha}$ in Eqs.(\ref{gEDPE},\ref{thetaEDPE},\ref{X_EDPE})
are denoted as $\bar{g}^{\rho}_{\alpha}$ in the four-body equations, while three-body energy-dependent
functions $\Theta(z)$ are denoted as $\bar{\tau}^{\rho}_{\alpha \beta}$ (four-body equations are written
down for  EDPA, i.e. with only one term  is taken in Eq.(\ref{X_EDPE}), $m=n=1$).

We tried different versions of numerical treatment of Eqs.(\ref{gHS},\ref{normHS}) and
Eqs.(\ref{gEDPE},\ref{thetaEDPE},\ref{X_EDPE}). 
The best result was obtained when the eigenequations
Eq.(\ref{gHS}) with normalization condition Eq.(\ref{normHS}) were solved using such number of
integration points, which is enough for exact calculations of the
corresponding binding energy of the three-body subsystem. For this sake 20 point is enough. 
The whole set of the eigenvalues and the corresponding eigenfunctions (in this case 20 eigenfunctions
for each of 20 eigenvalues) was evaluated.  We used ZGEEV subroutine from 
Intel oneAPI Math Kernel Library - Fortran, which computes
the eigenvalues and, optionally, the left and/or right eigenvectors for an n-by-n complex nonsymmetric matrix.
After that the first term in Eq.(\ref{X_EDPE}) with $m = n  = 1$ gives the correct binding energy 
(more exactly, $\Theta_{11}(z) {\longrightarrow \atop {z \to E_B}} \infty$) since
we strictly set the first eigenvalue $\lambda_1=1$  (of cause, when binding energy in the system exists).
We checked how the amplitudes converge, and found that, really, the first term already gives the result
close to the original value of $X_{\alpha \beta}(p,p';z) $ at the discrete set of momenta,
used in Eqs.(\ref{gHS},\ref{normHS}).

%%%%%%%%%%%%%%%%%%%%%%%%%%%%%%%%%%%%%%%%%%%%%%%%%%
\section{Four-body equations for the $\bar{K}NNN$ system}

\begin{figure}[htb]
\centering
\includegraphics[clip=true, trim=3.5cm 16cm 0cm 1.5cm, width=1.22\textwidth]{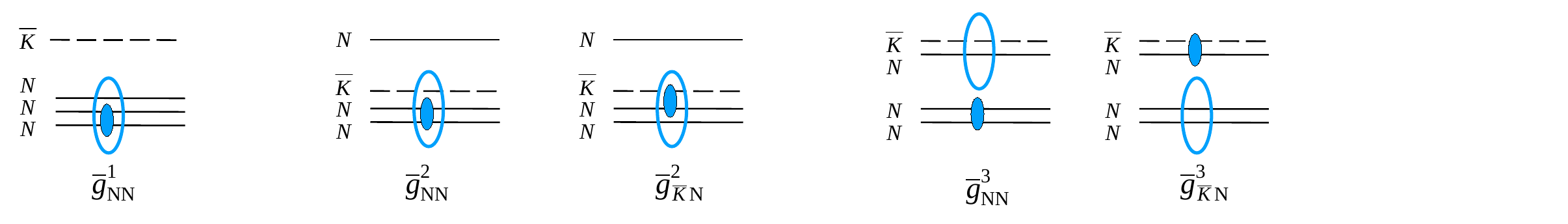}
\caption{States of the four-body system equations.} 
\label{g.fig}
\end{figure}

\begin{figure}[htb]
\centering
\includegraphics[clip=true, trim=2.1cm 9cm 0cm 1cm, width=1.05\textwidth]{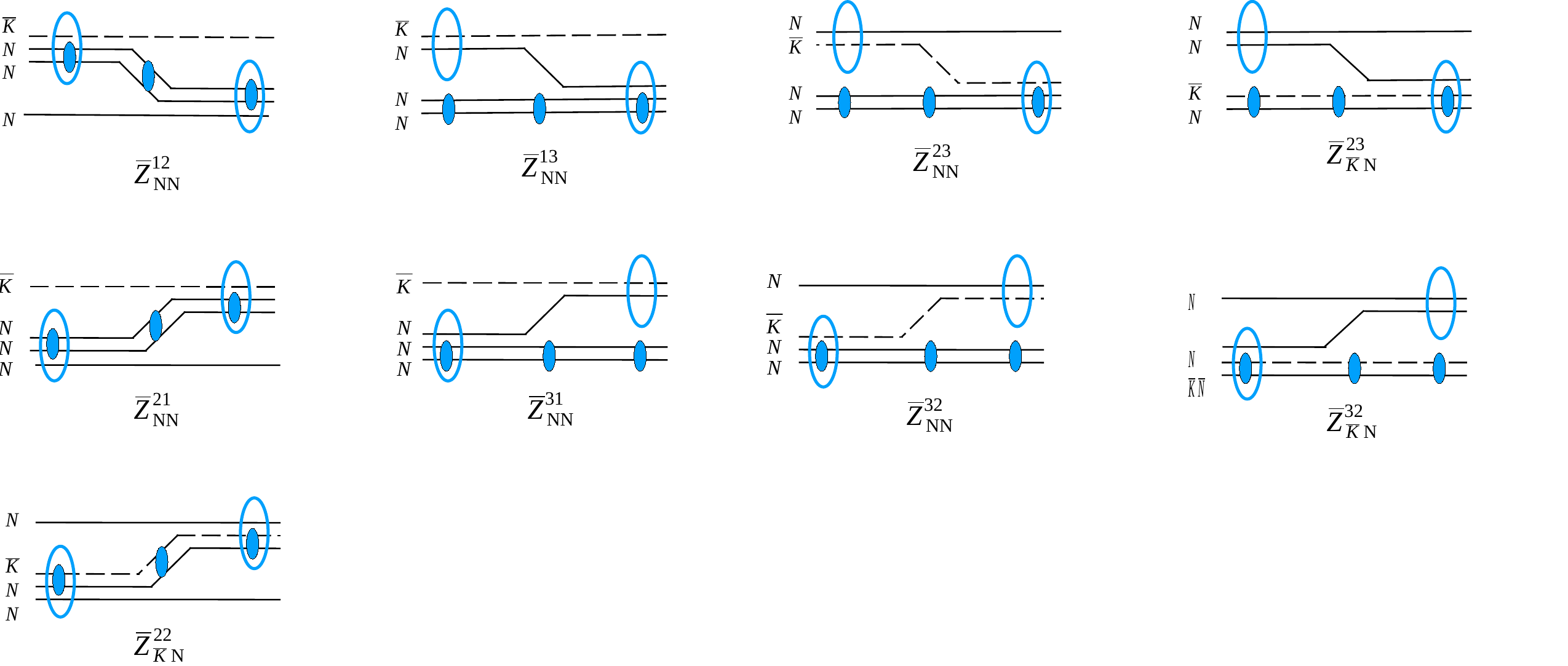}
\caption{Kernel functions $\bar{Z}^{\sigma \rho}_{\alpha}$ of the four-body system equations.} 
\label{Z.fig}
\end{figure}

Two types of partitions for a four-body system: $3+1$ and $2+2$, - for
the $\bar{K}NNN$ system are: $|\bar{K} + (NNN) \rangle$, $|N + (\bar{K}NN) \rangle$
and $|(\bar{K}N) + (NN) \rangle$. We started by writing down the system Eq.(\ref{4AGSsepVT}) for
$18$ channels considering three nucleons as nonidentical particles. Four-body
asymptotic states are denoted by ${}^\sigma_{\alpha}$ indices with $\sigma = 1,2,3$ 
stands for $|\bar{K} + (NNN) \rangle$, $|N + (\bar{K}NN) \rangle$ and $|(\bar{K}N) + (NN) \rangle$
partitions, correspondingly, and $\alpha =$ $N_i N_j$ or $\bar{K} N_i$ ($i,j = 1,2,3$, $i \ne j$) denotes
the pair in the two- or three-body subsystem:
\begin{eqnarray}
\nonumber
&{}&  \bar{g}^1_{N_iN_j}  :  |\bar{K} + (N_1 + N_2 N_3) \rangle, |\bar{K} + (N_2 + N_3 N_1) \rangle,
                      |\bar{K} + (N_3 + N_1 N_2) \rangle,          \\
\nonumber
&{}&  \bar{g}^2_{N_iN_j}:  |N_1 + (\bar{K} + N_2 N_3) \rangle, |N_2 + (\bar{K} + N_3 N_1) \rangle,
                      |N_3 + (\bar{K} + N_1 N_2) \rangle, \\
\label{channels}                      
&{}&  \bar{g}^2_{\bar{K}N_i}: |N_1 + (N_2 + \bar{K} N_3) \rangle, |N_2 + (N_3 + \bar{K} N_1) \rangle,
                      |N_3 + (N_1 + \bar{K} N_2) \rangle,\\
\nonumber                      
&{}&  \quad \quad \quad \! |N_1 + (N_3 + \bar{K} N_2) \rangle, |N_2 + (N_1 + \bar{K} N_3) \rangle,
                      |N_3 + (N_2 + \bar{K} N_1) \rangle,\\
\nonumber
&{}&  \bar{g}^3_{N_iN_j}:  |(N_2 N_3) + (\bar{K} + N_1) \rangle, |(N_3 N_1) + (\bar{K} + N_2) \rangle,
                   |(N_1 N_2) + (\bar{K} + N_3) \rangle, \\
\nonumber                  
&{}&  \bar{g}^3_{\bar{K}N_i}: |(\bar{K} N_1) + (N_2 + N_3) \rangle, |(\bar{K} N_2) + (N_3 + N_1) \rangle,
                      |(\bar{K} N_3) + (N_1 + N_2) \rangle
\end{eqnarray}
After antisymmetrization, necessary for a system with identical fermions, only five states, plotted
in Fig.\ref{g.fig}, remains.
%----------------------------------------------------------------------
%----------------------------------------------------------------------
The kernel functions $\bar{Z}^{\sigma \rho}_{\alpha}$  of the system of four-body equations
Eq.(\ref{4AGSsepVT}) can be seen in
Fig.\ref{Z.fig} ($\bar{Z}^{\sigma \rho}_{\alpha \beta}$ carrys only one bottom index due to
$\delta_{\alpha \beta}$ function in it's definition).
%----------------------------------------------------------------------
%----------------------------------------------------------------------

Looking for a quasi-bound state needs solution of the homogeneous system of equations,
which can be written in a matrix form
\begin{equation}
\label{4AGSfinal}
 \hat{X} = \hat{Z} \, \hat{\tau} \, \hat{X}
\end{equation}
with
\begin{equation}
\label{XZ5x5}
 \bar{X}^{\rho}_{\alpha} = \left(
 \begin{tabular}{c}
  $\bar{X}^{1}_{NN}$ \\
  $\bar{X}^{2}_{NN}$ \\
  $\bar{X}^{2}_{\bar{K}N}$ \\
  $\bar{X}^{3}_{NN}$ \\  
  $\bar{X}^{3}_{\bar{K}N}$ \\  
 \end{tabular}
 \right),
%=========================================
\qquad \qquad
%=========================================
 \bar{Z}^{\sigma \rho}_{\alpha} = \left(
 \begin{tabular}{ccccc}
  0 & $\bar{Z}^{12}_{NN}$ & 0 & $\bar{Z}^{13}_{NN}$ & 0 \\
  $\bar{Z}^{21}_{NN}$ & 0 & 0 & $\bar{Z}^{23}_{NN}$ & 0 \\
  0 & 0 & $\bar{Z}^{22}_{\bar{K}N}$ & 0 & $\bar{Z}^{23}_{\bar{K}N}$ \\
  $\bar{Z}^{31}_{NN}$ & $\bar{Z}^{32}_{NN}$ & 0 & 0 & 0 \\  
  0 & 0 & $\bar{Z}^{32}_{\bar{K}N}$ & 0 & 0 \\  
 \end{tabular}
 \right),
\end{equation}
%---------------------------------------------------------------------
\begin{equation}
\label{tau5x5}
 \bar{\tau}^{\rho}_{\alpha \beta} = \left(
 \begin{tabular}{ccccc}
  $\bar{\tau}^{1}_{NN,NN}$ & 0 & 0 & 0 & 0 \\
  0 & $\bar{\tau}^{2}_{NN,NN}$ & $\bar{\tau}^{2}_{NN,\bar{K}N}$ & 0 & 0 \\
  0 & $\bar{\tau}^{2}_{\bar{K}N,NN}$ & $\bar{\tau}^{2}_{\bar{K}N,\bar{K}N}$ & 0 & 0 \\
  0 & 0 & 0 & $\bar{\tau}^{3}_{NN,NN}$ & $\bar{\tau}^{3}_{NN,\bar{K}N}$ \\  
  0 & 0 & 0 & $\bar{\tau}^{3}_{\bar{K}N,NN}$ & $\bar{\tau}^{3}_{\bar{K}N,\bar{K}N}$ \\  
 \end{tabular}
 \right).
\end{equation}

However, our $\bar{K}N$ and $NN$ potentials, which we use for the $\bar{K}NNN$ system calculations,
are isospin- and spin-dependent interaction models, in addition $V_{NN}$ is a two-term
potential. Due to this elements of  the matrices $\bar{Z}^{\sigma \rho}_{\alpha}$ in Eq.(\ref{XZ5x5}) and
$\bar{\tau}^{\rho}_{\alpha \beta}$ in Eq.(\ref{tau5x5}), entering the antisymmetrized
equations Eq.(\ref{4AGSfinal}), are matrices themselves containing  elements with additional
indices 
$\bar{Z}^{\sigma \rho (m_3,n_3)}_{ {\alpha (m_2,n_2);} \atop {(i,ss')}}$
and $\bar{\tau}^{\sigma (m_3)}_{{\alpha \beta (m_2,m_2');} \atop {(i i',ss')}}$. 
Particular forms of nine elements-matrices $\bar{Z}^{\sigma \rho}_{\alpha}$ of
Eq.(\ref{XZ5x5}) and nine elements-matrices $\tau^{\rho}_{\alpha \beta}$ of Eq.(\ref{tau5x5}) 
are presented in the Appendix.
The additional indices $m_3, n_3$ denote number of a separable term of the
three-body or $2+2$ amplitudes 
(at the first step only one separable term was used for the three-body $\bar{K}NN$, $NNN$ and
2+2 $\bar{K}N + NN$ amplitudes in Eq. (\ref{X_EDPE}), so that $ m_3 = n_3= 1$).
Separable indices $m_2,n_2$ of the two-body subsystems (i.e. potentials) are
$m_2=1$ for $V_{\bar{K}N}$ and $m_2 = 1,2$ for $V_{NN}$. 
The remained indices $i, i'$ and $s, s'$ are two-body isospins and spins, correspondingly.

The unknown four-body amplitudes in Eq.(\ref{XZ5x5}) have a general form
$\bar{X}^{\sigma \rho (m_3,n_3)}_{\alpha \beta (m_2,n_2);(i i', s s')}$. Finally, the system
to be solved consists of $18$ coupled equations.

%%%%%%%%%%%%%%%%%%%%%%%%%%%%%%%%%%%%%%%%%%%%%%%%%%%
\section{Two-body interactions and three-body subsystems}

%---
\subsection{Two-body input: $\bar{K}N$ and $NN$ potentials}

Both $\bar{K}N$ and $NN$ potentials, which we used, are separable isospin- and spin-dependent ones in
$s$-wave. Three our separable antikaon-nucleon potentials were constructed for our three-body calculations
of the $\bar{K}NN$ and $\bar{K} \bar{K} N$ systems \cite{review}. They are:  two phenomenological potentials with
coupled $\bar{K}N - \pi \Sigma$ channels, having one- $V_{\bar{K}N}^{\rm 1,SIDD}$ or two-pole 
$V_{\bar{K}N}^{\rm 2,SIDD}$ structure of the $\Lambda(1405)$ resonance~\cite{NPA890} and
a chirally motivated model $V_{\bar{K}N}^{\rm Chiral}$ with coupled $\bar{K}N - \pi \Sigma - \pi \Lambda$
channels and the two-pole structure~\cite{PRC90-I}. All three potentials describe low-energy $K^- p$ scattering,
namely: elastic $K^- p \to K^- p$ and inelastic $K^- p \to MB$ cross-sections and threshold branching ratios $\gamma, R_c, R_n$.
They also reproduce $1s$ level shift of kaonic hydrogen caused by the strong $\bar{K}N$ interaction
in comparison to the pure Coulomb level, measured by SIDDHARTA experiment~\cite{SIDD}: 
$\Delta_{1s}^{SIDD} = -283 \pm 36 \pm 6$ eV, and it's width $\Gamma_{1s}^{SIDD} = 541 \pm 89 \pm 22$ eV. All
the experimental data are described by three our potentials with equally high accuracy. In addition,
elastic $\pi \Sigma$ cross-sections with isospin $I_{\pi \Sigma} = 0$, provided by all three potentials,
have a bump in a region of the $\Lambda(1405)$ resonance (according to PDG~\cite{PDG}:
$M_{\Lambda(1405)}^{PDG} = 1405.1^{+1.3}_{-1.0}$ MeV,
$\Gamma_{\Lambda(1405)}^{PDG} = 50.5 \pm 2.0$ MeV).
The poles corresponding to the $\Lambda(1405)$ resonance are situated at
\begin{eqnarray}
\label{zLambda1405}
z_{\Lambda(1405)-1}^{\rm 1,SIDD} &=& 1426 - i \, 48 {\, \rm MeV} \\
z_{\Lambda(1405)-1}^{\rm 2,SIDD} &=& 1414 - i \, 58 {\, \rm MeV}, \quad
z_{\Lambda(1405)-2}^{\rm 2,SIDD} = 1386 - i \, 104 {\, \rm MeV}
\end{eqnarray}
for the phenomenological potentials with one- and two-pole structure, correspondingly~\cite{PRC90-I}, and at
\begin{equation}
z_{\Lambda(1405)-1}^{\rm Chiral} = 1417 - i \, 33 {\, \rm MeV}, \quad
 z_{\Lambda(1405)-2}^{\rm Chiral} = 1406 - i \, 89 {\, \rm MeV}
\end{equation}
for the chirally motivated potential~\cite{PRC90-II}.

These three antikaon-nucleon potentials with coupled $\bar{K}N - \pi \Sigma$ channels were used 
in their original form in three-body AGS equations \cite{review} with coupled $\bar{K}NN - \pi \Sigma N$
three-body channels. By this the channel coupling was taken into account in a direct way. The four-body GS
equations Eq.(\ref{4AGSsepVT}) are too complicated for introducing additional particle channels and performing
coupled-channel calculations. Due to this we used the exact optical versions of our $\bar{K}N$ potentials
\cite{PRC_myKd}, which are one-channel $V_{\bar{K}N(-\pi \Sigma-\pi \Lambda)}$. They have exactly
the same elastic part as the potential with coupled channels, while all in-elasticity is taken into account in
an energy-dependent imaginary part of the potential. It was demonstrated in our three-body calculations
\cite{PRC_myKd,KdQBS}, that such exact optical potentials give quite accurate results for $K^- d$
scattering length or quasi-bound state position and width of the $K^- pp$ system in comparison with
the results obtained with
the coupled-channel form of interaction models. Due to this we assume that it is a good
approximation for the four-body calculations as well.

Nucleon-nucleon potentials $V^{\rm TSA-A}_{NN}$ and $V^{\rm TSA-B}_{NN}$ used in our three-body calculations \cite{review} were used here together
with the new version of the two-term separable $NN$ potential $V^{\rm TSN}_{NN}$, described in \cite{KdQBS}. 
All three potentials reproduce Argonne v18 $NN$ phase shifts at low energies up to $500$ MeV with change
of sign, which means they are repulsive at short distances. They give proper singlet
and triplet $NN$ scattering lengths and deuteron binding energy. The new nucleon-nucleon potential 
$V^{\rm TSN}_{NN}$ reproduces Argonne v18 $NN$ phase shifts
of $pp$ scattering slightly better than the previously used ones, its parameters are more natural then those
of the older $V_{NN}^{\rm TSA}$.

In principle, Coulomb interaction also should be included in the equations, but we are interested in the
quasi-bound state caused mainly by strong potentials. We assume that in such a  state Coulomb interaction
plays a minor role and can be neglected.

%---
\subsection{$3+1$ and $2+2$ partitions}

We investigated $\bar{K}NNN$ system with the lowest value of the four-body isospin $I^{(4)}=0$ and
spin $S^{(4)}$ one half, which can be denoted as $K^- ppn$ or $\bar{K}^0 nnp$. The total angular
momentum is zero, all two-body interactions are $s$-wave ones. For the $\bar{K}NNN$ system
with these quantum numbers the following three-body subsystems contribute:
\begin{itemize}
 \item $\bar{K}NN$ with  isospin $I^{(3)} = 1/2$ and spin $S^{(3)} = 0$ ($K^- pp$) or spin $S^{(3)} = 1$ ($K^- np$)
 \item $NNN$ with isospin $I^{(3)} = 1/2$ and spin $S^{(3)} = 1/2$ ($^{3}$H or $^{3}$He)
\end{itemize}
together with the $2+2$ partition
\begin{itemize}
 \item $\bar{K}N+NN$ with  isospin $I^{(4)} = 0$ and spin $S^{(4)} = 1/2$.
\end{itemize}
The corresponding three-body amplitudes is not an input, they are calculated during the four-body calculations.

%----------------------------------------------------------------
\begin{table}[ht]
\caption{Dependence of the binding energy $B_{K^-pp}^{\rm Opt}$ (MeV) and width
$\Gamma_{K^- pp}^{\rm Opt}$ (MeV)
of the quasi-bound state in the $K^- pp - \bar{K}^0 np$ subsystem  ($\bar{K}NN, S^{(3)}=0$)  on three
$\bar{K}N$ and three $NN$ interaction models. Calculations were performed using the exact optical
$\bar{K}N$ potentials.
The binding energy is counted from the threshold energy of the
$K^-pp$ system  $z_{th, {\rm K^- pp}} = m_{\bar{K}}+2 \, m_N$ $=2373.485$ MeV.}
\label{polesKpp.tab}
\begin{center}
\begin{tabular}{ccccccc}
\hline \noalign{\smallskip}
  & \multicolumn{2}{c}{$V_{NN}^{\rm TSA-A}$} 
    & \multicolumn{2}{c}{$V_{NN}^{\rm TSA-B}$}  
     & \multicolumn{2}{c}{$V_{NN}^{\rm TSN}$}  \\
  & $B_{K^-pp}^{\rm Opt}$ \; &  $\Gamma_{K^- pp}^{\rm Opt}$ \;  & $B_{K^-pp}^{\rm Opt}$ \;  & $\Gamma_{K^- pp}^{\rm Opt}$ \;  
        & $B_{K^-pp}^{\rm Opt}$ \;  & $\Gamma_{K^- pp}^{\rm Opt}$  \\
\noalign{\smallskip} \hline \noalign{\smallskip}
  $V_{\bar{K}N}^{\rm 1,SIDD}$   & $55.4$  & $60.9$ & $54.3$ & $60.8$ & $53.3$ & $64.7$   \\
  $V_{\bar{K}N}^{\rm 2,SIDD}$ & $48.2$  & $46.2$ & $47.5$ & $45.9$ & $46.7$ & $48.4$  \\
  $V_{\bar{K}N}^{\rm Chiral}$ & $31.9$  & $42.2$ & $33.2$ & $48.7$ & $29.9$ & $48.2$  \\
 \noalign{\smallskip} \hline
\end{tabular}
\end{center}
\end{table}
%----------------------------------------------------------------

The three-body $\bar{K}NN$ system with different quantum numbers was studied in our previous works.
In particular, quasi-bound state pole positions and widths in the $K^- pp$ system ($\bar{K}NN$ with isospin
$I^{(3)} = 1/2$ and spin $S^{(3)} = 0$) were calculated in~\cite{PRC90-II} with older
$V_{NN}^{\rm TSA-B}$
nucleon-nucleon potential. Recently the calculations were repeated with the new $V_{NN}^{\rm TSN}$
\cite{KdQBS}, the results can be found in Table 4 of the paper.  Binding energies $B_{K^- pp, \rm Opt}$
and widths $\Gamma_{K^- pp, \rm Opt}$ calculated using the exact optical potentials are shown in
Table~\ref{polesKpp.tab} for three antikaon-nucleon $V_{\bar{K}N}^{\rm 1,SIDD}$,
$V_{\bar{K}N}^{\rm 2,SIDD}$, $V_{\bar{K}N}^{\rm Chiral}$ and three nucleon-nucleon 
$V_{NN}^{\rm TSA-A}$, $V_{NN}^{\rm TSA-B}$, $V_{NN}^{\rm TSN}$ potentials.
It is seen that the new $NN$ potential changed quasi-bound state positions in $K^- pp$ system by
few MeV.

%----------------------------------------------------------------
\begin{table}[ht]
\caption{Dependence of the binding energy $B_{K^-np}^{\rm Opt}$ (MeV) and 
width $\Gamma_{K^- np}^{\rm Opt}$ (MeV)
of the quasi-bound state in the $K^- np - \bar{K}^0 nn$ subsystem ($\bar{K}NN, S^{(3)}=1$)  on 
hree $\bar{K}N$ and three $NN$ interaction models. Calculations were performed using the exact optical
$\bar{K}N$ potentials. The energy is counted from the $K^- d$ threshold
$z_{th, {\rm K^- d}} = m_{\bar{K}} + 2 m_N + E_{\rm deu}$ $=2371.26$ MeV.}
\label{polesKnp.tab}
\begin{center}
\begin{tabular}{ccccccc}
\hline  \noalign{\smallskip}
  & \multicolumn{2}{c}{$V_{NN}^{\rm TSA-A}$} 
    & \multicolumn{2}{c}{$V_{NN}^{\rm TSA-B}$}  
     & \multicolumn{2}{c}{$V_{NN}^{\rm TSN}$}  \\
  & $B_{K^-np}^{\rm Opt}$ \; &  $\Gamma_{K^- np}^{\rm Opt}$ \;  & $B_{K^-np}^{\rm Opt}$ \;  & $\Gamma_{K^- np}^{\rm Opt}$ \;  
        & $B_{K^-np}^{\rm Opt}$ \;  & $\Gamma_{K^- np}^{\rm Opt}$  \\
\noalign{\smallskip} \hline \noalign{\smallskip}
  $V_{\bar{K}N}^{\rm 1,SIDD}$   & $1.6$  & $70.1$ & $0.8$ & $67.6$ & $1.9$ & $68.7$   \\
  $V_{\bar{K}N}^{\rm 2,SIDD}$ & $5.2$  & $63.7$ & $5.0$ & $61.4$ & $5.6$ & $62.7$  \\
  $V_{\bar{K}N}^{\rm Chiral}$ & $2.6$  & $46.4$ & $2.4$ & $53.1$ & $2.3$ & $45.5$  \\
 \noalign{\smallskip} \hline 
\end{tabular}
\end{center}
\end{table}
%----------------------------------------------------------------

No quasi-bound states similar to that one in $K^- pp$ were found in 
 the $K^- np$ system, which is $\bar{K}NN$ with isospin $I^{(3)} = 1/2$ and spin $S^{(3)} = 1$
in our previous calculations \cite{PRC90-I}. However, new nucleon-nucleon potential $V^{\rm TSN}_{NN}$
changed the picture: the quasi-bound state caused purely by strong
interactions can exist in the $K^-np$ system \cite{PRC_myKd} in addition to the atomic kaonic deuterium.
The binding energies $B_{K^- np,\rm Opt}$ and width $\Gamma_{K^- np, \rm Opt}$ evaluated with
the exact optical versions of the
$V_{\bar{K}N}^{\rm 1,SIDD}$, $V_{\bar{K}N}^{\rm 2,SIDD}$, $V_{\bar{K}N}^{\rm Chiral}$ 
potentials and three nucleon-nucleon 
$V_{NN}^{\rm TSA-A}$, $V_{NN}^{\rm TSA-B}$, $V_{NN}^{\rm TSN}$ potentials
can be seen in Table~\ref{polesKnp.tab}.
The binding energy of the state is so close to the $K^- d$ threshold, that relatively weak $NN$ interaction,
playing in the case of strong quasi-bound state secondary role, can resolve the question of existence
of the state.

The binding energies and widths of the quasi-bound states of the $\bar{K}NN$ systems with spin zero
and one were calculated using the three-body AGS equations.
Details of three-body calculations can be found in \cite{PRC_myKd,PRC_1stKpp}.
The codes for numerical solution of the three-body AGS equations for the $\bar{K}NN$ systems were
than modified in order to construct separable versions of the three-body amplitudes, as described in
Section \ref{separab.sec}.

%----------------------------------------------------------------
\begin{table}[t]
\caption{Dependence of the binding energy $B_{NNN}$ (MeV) and width $\Gamma_{NNN}$ (MeV)
of the quasi-bound state in the $NNN$ subsystem ($S^{(3)}=1/2$) on three $NN$ models. 
The binding energy is counted from the threshold energy of the
$NNN$ system  $z_{th, {\rm NNN}} = 3 \, m_N$ $= 2816.76$ MeV.}
\label{polesNNN.tab}
\begin{center}
\begin{tabular}{ccc}
\hline  \noalign{\smallskip}
  $V_{NN}^{\rm TSA-A}$    \quad &   $V_{NN}^{\rm TSA-B}$ \quad   &  $V_{NN}^{\rm TSN}$ \\
 \noalign{\smallskip} \hline \noalign{\smallskip}
  $9.03$    &    $9.04$     & $9.52$    \\
 \noalign{\smallskip} \hline 
\end{tabular}
\end{center}
\end{table}
%----------------------------------------------------------------

The three-body AGS equations Eq.(\ref{3AGSsep}) were written and numerically solved for the three-nucleon
system $NNN$ with three $NN$ potentials $V_{NN}^{\rm TSA-A}$, $V_{NN}^{\rm TSA-B}$,
$V_{NN}^{\rm TSN}$ as an input. The calculated binding energies of the system are
equal for both ${}^3$H and ${}^3$He nuclei since Coulomb interaction was not taken
into account. They are presented in Table~\ref{polesNNN.tab}.
The resulting energies are larger than known values for binding energy of triton $8.4820$ MeV and
helium-3 $7.7181$ MeV nucleus. Such overestimation is typical for separable $NN$ potentials.
The numerical code was afterwards
changed for construction of separable version of the $NNN$ amplitude.
%----------------------------------------------------------------
\begin{table}[h]
\caption{Dependence of the binding energy $B_{K^-p+np}^{\rm Opt}$ (MeV) and 
width $\Gamma_{K^- p+np}^{\rm Opt}$ (MeV)
of the quasi-bound state in the $K^- p+np$ partition ($\bar{K}N + NN, S^{(4)}=1/2$) on three $\bar{K}N$
and three $NN$ interaction models. Calculations were performed using the exact optical
$\bar{K}N$ potentials.
The binding energy is counted from the threshold energy of the
$K^-npp$ system  $z_{th, {\rm K^- npp}} = m_{\bar{K}}+3 \, m_N$ $=3312.405$ MeV.}
\label{polesKp+np.tab}
\begin{center}
\begin{tabular}{ccccccc}
\hline \noalign{\smallskip}
  & \multicolumn{2}{c}{$V_{NN}^{\rm TSA-A}$} 
    & \multicolumn{2}{c}{$V_{NN}^{\rm TSA-B}$}  
     & \multicolumn{2}{c}{$V_{NN}^{\rm TSN}$}  \\
  & $B_{K^-p+np}^{\rm Opt}$ \; &  $\Gamma_{K^- p+np}^{\rm Opt}$ \;  & $B_{K^-p+np}^{\rm Opt}$ \;  
& $\Gamma_{K^- p+np}^{\rm Opt}$ \;  
        & $B_{K^-p+np}^{\rm Opt}$ \;  & $\Gamma_{K^- p+np}^{\rm Opt}$  \\
\noalign{\smallskip} \hline \noalign{\smallskip}
  $V_{\bar{K}N}^{\rm 1,SIDD}$   & $20.0$  & $83.6$ & $19.1$ & $81.1$ & $20.3$ & $82.2$   \\
  $V_{\bar{K}N}^{\rm 2,SIDD}$ & $25.4$  & $71.3$ & $24.0$ & $70.8$ & $25.1$ & $70.4$  \\
  $V_{\bar{K}N}^{\rm Chiral}$ & $18.8$  & $54.5$ & $20.9$ & $58.5$ & $17.9$ & $53.7$  \\
 \noalign{\smallskip} \hline
\end{tabular}
\end{center}
\end{table}
%----------------------------------------------------------------

Finally, the partition of the $2+2$ type $\bar{K}N + NN$ is a system with two non-interacting pairs of particles.
It is a special part for the four-body Faddeev-type equations and not a simple sum of two-body $\bar{K}N$
and $NN$ amplitudes. $\bar{K}N + NN$ partition is described by a three-body system of AGS equations, therefore, it is a "three-body" amplitude.
Only $\bar{K}N + NN$ partition with spin one $NN$ pair has a quasi-bound state, the binding
energies and widths are shown in Table~\ref{polesKp+np.tab}.
The separable version of the corresponding $\bar{K}N + NN$ amplitude was constructed in similar way 
as in the case of $\bar{K}NN$ and $NNN$ three-body subsystems.

%%%%%%%%%%%%%%%%%%%%%%%%%%%%%%%%%%%%%%%%%%%%%%%%
\section{Results and discussion}

We solved the four-body GS equations for the $K^- ppn - \bar{K}^0 nnp$ system with
the antikaon-nucleon and nucleon-nucleon potentials described above. As the first step we
used only one separable term in 
separable representation of the three-body $\bar{K}NN$, $NNN$ and "three-body"
$\bar{K}N+NN$ amplitudes Eq.(\ref{X_EDPE}), so that we used EDPA method. The binding
energies $B_{K^- ppn}$ and widths $\Gamma_{K^- ppn}$ of the $\bar{K}NNN$ system
evaluated using three exact optical versions of the antikaon-nucleon potentials and three nucleon-
nucleon interaction models are presented in Table \ref{polesKNNN.tab}.

It is seen that the binding energy $B_{K^-ppn}$ and width $\Gamma_{K^- ppn}$ of the four-body
quasi-body state strongly depend on the model of antikaon-nucleon interaction. In fact, it is
a property of all three-body $\bar{K}NN$ and $\bar{K} \bar{K}N$ systems as well.
Phenomenological $\bar{K}N (- \pi \Sigma)$ potentials give comparable binding energies,
while the chirally motivated potential led to much more shallow state. As for the widths, the
models of antikaon-nucleon interaction having two-pole structure of $\Lambda(1405)$ resonance,
which are the phenomenological $V_{\bar{K}N}^{\rm 2,SIDD}$ and chiral $V_{\bar{K}N}^{\rm Chiral}$,
have very close widths, while one-pole $V_{\bar{K}N}^{\rm 1,SIDD}$ potential lead to much
larger width.

%----------------------------------------------------------------
\begin{table}[ht]
\caption{Dependence of the binding energy $B_{K^-ppn}$ (MeV) and width $\Gamma_{K^- ppn}$ (MeV)
of the quasi-bound state in the $K^- ppn - \bar{K}^0 nnp$ system on three $\bar{K}N$ and $NN$
interaction models.  }
\label{polesKNNN.tab}
\begin{center}
\begin{tabular}{ccccccccc}
\hline  \noalign{\smallskip}
  & \multicolumn{2}{c}{$V_{NN}^{\rm TSA-A}$} 
    & \multicolumn{2}{c}{$V_{NN}^{\rm TSA-B}$}  
     & \multicolumn{2}{c}{$V_{NN}^{\rm TSN}$}  
     & \multicolumn{2}{c}{\rm Other results} \\
  & $B_{K^-ppn}$ \; &  $\Gamma_{K^- ppn}$ \;  & $B_{K^-ppn}$ \;  & $\Gamma_{K^- ppn}$ \;  
        & $B_{K^-ppn}$ \;  & $\Gamma_{K^- ppn}$ \; & $B_{K^-ppn}$ \;  & $\Gamma_{K^- ppn}$ \\
\noalign{\smallskip} \hline \noalign{\smallskip}
   $V_{\bar{K}N}^{\rm 1,SIDD}$ & $52.0$  & $50.4$ & $50.3$ & $49.6$ & $51.2$ & $50.8$ & & \\
  $V_{\bar{K}N}^{\rm 2,SIDD}$ & $47.0$  & $39.6$ & $46.4$ & $38.2$ & $46.4$ & $39.9$ & & \\
  $V_{\bar{K}N}^{\rm Chiral}$   & $32.6$  & $39.7$ & $34.5$ & $50.9$ & $30.5$ & $42.8$ & &  \\
 \noalign{\smallskip} \hline
   AY \cite{AY1}          & & & & & & & 108.0 & 20.0 \\
   BGL \cite{BGL}       & & & & & & & 29.3 & 32.9 \\
   OHHMH \cite{OHHMH}: & & & & & & &  &   \\
    $V_{\bar{K}N}^{\rm Kyoto-I}$ & & & & & & & 45.3 & 25.5  \\
    $V_{\bar{K}N}^{\rm Kyoto-II}$ & & & & & & & 49.7 & 69.4  \\
    $V_{\bar{K}N}^{\rm AY}$ & & & & & & & 72.6 & 78.6  \\
   ME \cite{ir}:              & & & & & & &  &  \\
    $V_{\bar{K}N}^{\rm 1,SIDD}$              & & & & & & & 73.5 & 22.0 \\
    $V_{\bar{K}N}^{\rm 2,SIDD}$               & & & & & & & 58.5 & 27.0 \\
    $V_{\bar{K}N}^{\rm IKS \, chiral}$               & & & & & & & 41.4 & 31.5 \\
 \noalign{\smallskip} \hline
\end{tabular}
\end{center}
\end{table}
%----------------------------------------------------------------

As for the dependence of the obtained results on the models of $NN$ interactions, it seem they
play more visible role than in the case of $\bar{K}NN$ system. Nucleon-nucleon
interaction plays visible role together with the chirally motivated antikaon nucleon potential, while the
results obtained with phenomenological $\bar{K}N$ interaction models depend on the $V_{NN}$
slightly.

Comparing the four-body binding energies and widths in Table~\ref{polesKNNN.tab} with those 
obtained for the three-body spin zero $\bar{K}NN$ system usually denoted as $K^- pp$
Table~\ref{polesKpp.tab}, it is seen that the binding energies remained almost the same or
become slightly smaller after adding one neutron, while the widths become much smaller.
So that the additional to $K^- pp$ neutron changes the binding energy of the system slightly, but
it "tightenes" the quasi-bound state.

Obviously, the four-body binding energies are much larger and the corresponding widths are
much smaller than those for the other, spin one state of the $\bar{K}NN$ state ($K^- np$).

Results for the binding energies and widths of other authors are also presented in
Table~\ref{polesKNNN.tab}. The largest binding energy and
smaller width were predicted in \cite{AY1} by using many-body G-matrix formalism
for the few-body system and an antikaon-nucleon model, which does not reproduce actual experimental
data on $K^- p$ scattering and kaonic hydrogen. 

The authors of variational  calculation \cite{BGL} used an energy dependent chiral $\bar{K}N$ potential,
which, however,  was calculated at set of fixed energies. Also ithe maginary part of the quasi-bound state
was calculated approximately. The result of \cite{BGL}  is not far from our binding energy and width
evaluated with our chirally motivated antikaon-nucleon potential, taken into account exactly. 

Variational calculations were also performed in  \cite{OHHMH} with energy dependent Kyoto and energy
independent AY \cite{AYVKN} $\bar{K}N$ potentials. The authors also had to fix the antikaon-nucleon energy in
the Kyoto potential, and they did it in two ways, denoting the corresponding models as Kyoto-I
and Kyoto-II. Huge difference between the widths obtained with these two Kyoto potentials shows
that chosen procedure is not quite reliable. Besides, the binding energy and width of the four-body
quasi-bound state, calculated in \cite{OHHMH} with AY potnetial \cite{AYVKN}, differs drastically from
the original AY results \cite{AY1}.

The most intriguing is the difference between our results and those of \cite{ir} since the authors
solved the same four-body Faddeev-type equations, moreover, they used our phenomenological 
$V_{\bar{K}N}^{\rm 1,SIDD}$ and $V_{\bar{K}N}^{\rm 2,SIDD}$ potentials.
Nucleon-nucleon PEST $V_{NN}$ potential, which is
a separable version of Paris nucleon-nucleon potential, was used there. But we do not think, 
that different from our's $NN$ potential could
change the results so drastically, by $\sim 20$ MeV, for both: binding energy (make it much larger) and
width (make it much smaller). 

One of the possible reasons of differences could follow from the way of separable version of the three-body
and
$2+2$ amplitudes construction. The formulas for the EDPE energy dependent form-factors and propagators
in \cite{ir} differ from our Eqs.(\ref{gEDPE},\ref{thetaEDPE}) by $1/\lambda$ factor on the right-hand
side for the first and by $m \leftrightarrow n$ in the second formula. We checked the formulas
by comparing the approximate $Z_{\alpha \beta}(p,p';z)$ at fixed momenta and energy with the
original ones, and obtained agreement up to $\sim 10$ significant digits while the whole number
of separable terms (which is 20 in our case) is used.

Besides, setting the binding energy  for the $(\bar{K}N)_{I=0} + NN$ to the energy of the quasi-bound
$\bar{K}N$ state ($\Lambda(1405)$ resonance), made in \cite{ir}, is not quite understandable. The matter
is the total four-body isospin $I^{(4)}=0$ means that the $NN$ subsystem should
also have $I_{NN}=0$, which is a deuteron. Due to this, the total binding energy of this partition
should include both two-body energies: $B_{\Lambda(1405)}$ and $B_{\rm deu}$. In fact, the energy of
the $\bar{K}N + NN$ "three-body"
system with isospins of both pairs equal to zero differs from simple sum of the two two-body energies,
see Table \ref{polesKp+np.tab}, it comes from solution of a three-body equation. 
As for another state of the
$\bar{K}N + NN$ partition, it does not have bound states since $\bar{K}N$ and $NN$ in this case have
two-body isospin $I^{(2)}=1$ and none of them is bound. Setting the energy in this case to deuteron
binding energy, as is done in \cite{ir}, is quite strange. 

%%%
Our very preliminary results for the pole position of the quasi-bound state in the $K^- ppn - \bar{K}^0 nnp$
system calculated with two phenomenological potentials were published in \cite{EFBGuilfordProc}. They
differ from those presented in Table~\ref{polesKNNN.tab} drastically: binding energies are much larger
while the width are smaller. The reason is the different procedure of the separable representation of
the three- and "three"-body amplitudes. In contrast to the present usage of the Intel MKL library subroutine,
we used some hand made procedure for eigenvalues and eigenfunctions evaluation. Such difference in
the procedure also can explain differences between our present results and those of \cite{ir}. 

Finally, the binding energy the three-body $K^- pp$ quasi-bound state evaluated in \cite{ir}  with our
phenomenological $V_{\bar{K}N}^{\rm 1,SIDD}$ potential differ from our result by $\sim 6$ MeV and
almost coinsides with the energy calculated with $V_{\bar{K}N}^{\rm 2,SIDD}$. Our three-body 
binding energies are sufficiently different for every of the three nucleon-nucleon potentials,
see Table \ref{polesKpp.tab}. Keeping this closeness of the three-body results in mind, it is hard to
understand large difference ($15$ MeV) between the corresponding four-body results in \cite{ir}.

\section{Conclusion}

The four-body Faddeev-type GS equations for search of the quasi-bound state in the $\bar{K}NNN$
system were written down and solved. The binding energies $B^{\rm Chiral}_{K^-ppn} \sim 30.5 - 34.5$ MeV
obtained with chirally motivated and $B^{\rm SIDD}_{K^-ppn} \sim 46.4 - 52.0$ MeV obtained with
phenomenological antikaon-nucleon potentials are close to those for the $K^- pp$ system, calculated
with the same $V_{\bar{K}N}$ and $V_{NN}$ potentials.
The widths of the four-body states $\Gamma_{K^- ppn} \sim 38.2 - 50.9$ MeV
are smaller than the three-body widths of $K^- pp$. Therefore, the neutron 
added to the $K^- pp$ system slightly influences the binding, but tightens the system.

The four-body binding energy and width of the $\bar{K}NNN$ system strongly depend on $\bar{K}N$
potential and noticeably - on $NN$ potential, especially together with the chirally motivated 
antikaon-nucleon interaction model.

\section*{Acknowledgments}
The work was supported by GACR grant 19-19640S.

\appendix*
\section{Elements of $\bar{Z}^{\sigma \rho}_{\alpha}$ and $\bar{\tau}^{\rho}_{\alpha \beta}$ matrices}

After antisymmetrization  we come to the system Eq.(\ref{4AGSfinal}) of $18$ coupled equations.
Elements of  the matrices $\bar{Z}^{\sigma \rho}_{\alpha}$ in Eq. (\ref{XZ5x5}) and
$\bar{\tau}^{\rho}_{\alpha \beta}$ in Eq.(\ref{tau5x5}) are
matrices themselves with elements $\bar{Z}^{\sigma \rho (m_3,n_3)}_{ {\alpha (m_2,n_2);} \atop {(i,ss')}}$
and $\bar{\tau}^{\sigma (m_3)}_{{\alpha \beta (m_2,m_2');} \atop {(i i',ss')}}$.  
Indices $m_3, n_3$ denote number of a separable term of the three-body or $2+2$ amplitudes
($ m_3 = n_3= 1$), while $m_2,n_2$ are separable indices
of the potentials: $m_2=1$ for $V_{\bar{K}N}$ and $m_2 = 1,2$ for $V_{NN}$. 
 Two-body isospins and spins are denoted $i, i'$ and $s, s'$, correspondingly.

The kernel matrices $\bar{Z}^{\sigma \rho}_{\alpha}$ consists of six elements with $\alpha = NN$:
%---------------------------------------------------------------------------------------------------------------------------------------------------------------
\begin{equation}
\label{Z12NN}
\bar{Z}^{12}_{NN} = \left(
\begin{tabular}{cccc}
 $\bar{Z}^{12(1,1)}_{{NN(1,1);} \atop {(0,11)}}$ & 0 & $\bar{Z}^{12(1,1)}_{{NN(1,2);} \atop {(0,11)}}$ & 0 \\
 
 0 & $\bar{Z}^{12(1,1)}_{{NN(1,1);} \atop {(1,00)}}$ & 0 & $\bar{Z}^{12(1,1)}_{{NN(1,2);} \atop {(1,00)}}$  \\

 $\bar{Z}^{12(1,1)}_{{NN(2,1);} \atop {(0,11)}}$ & 0 & $\bar{Z}^{12(1,1)}_{{NN(2,2);} \atop {(0,11)}}$ & 0   \\

 0 & $\bar{Z}^{12(1,1)}_{{NN(2,1);} \atop {(1,00)}}$ & 0 & $\bar{Z}^{12(1,1)}_{{NN(2,2);} \atop {(1,00)}}$ 
\end{tabular}
\right)
\end{equation}
%---------------------------------------------------------------------------------------------------------------------------------------------------------------
\begin{equation}
\label{Z13NN}
\bar{Z}^{13}_{NN} = \left(
\begin{tabular}{cccc}
$\bar{Z}^{13(1,1)}_{{NN(1,1);} \atop {(0,11)}}$ & 0 & $\bar{Z}^{13(1,1)}_{{NN(1,2);} \atop {(0,11)}}$ & 0  \\

 0 & $\bar{Z}^{13(1,1)}_{{NN(1,1);} \atop {(1,00)}}$ & 0 & $\bar{Z}^{13(1,1)}_{{NN(1,2);} \atop {(1,00)}}$   \\

 $\bar{Z}^{13(1,1)}_{{NN(2,1);} \atop {(0,11)}}$ & 0 & $\bar{Z}^{13(1,1)}_{{NN(2,2);} \atop {(0,11)}}$ & 0   \\

 0 & $\bar{Z}^{13(1,1)}_{{NN(2,1);} \atop {(1,00)}}$ & 0 & $\bar{Z}^{13(1,1)}_{{NN(2,2);} \atop {(1,00)}}$  
\end{tabular}
\right)
\end{equation}
%---------------------------------------------------------------------------------------------------------------------------------------------------------------
\begin{equation}
\label{Z21NN}
\bar{Z}^{21}_{NN} = \left(
\begin{tabular}{cccc}
$\bar{Z}^{21(1,1)}_{{NN(1,1);} \atop {(0,11)}}$ & 0 & $\bar{Z}^{21(1,1)}_{{NN(1,2);} \atop {(0,11)}}$ & 0 \\

 0 & $\bar{Z}^{21(1,1)}_{{NN(1,1);} \atop {(1,00)}}$ & 0 & $\bar{Z}^{21(1,1)}_{{NN(1,2);} \atop {(1,00)}}$ \\

$\bar{Z}^{21(1,1)}_{{NN(2,1);} \atop {(0,11)}}$ & 0 & $\bar{Z}^{21(1,1)}_{{NN(2,2);} \atop {(0,11)}}$ & 0  \\

0 & $\bar{Z}^{21(1,1)}_{{NN(2,1);} \atop {(1,00)}}$ & 0 & $\bar{Z}^{21(1,1)}_{{NN(2,2);} \atop {(1,00)}}$ 
\end{tabular}
\right)
\end{equation}
%---------------------------------------------------------------------------------------------------------------------------------------------------------------
\begin{equation}
\label{Z23NN}
\bar{Z}^{23}_{NN} = \left(
\begin{tabular}{cccc}

$\bar{Z}^{23(1,1)}_{{NN(1,1);} \atop {(0,11)}}$ & 0 & $\bar{Z}^{23(1,1)}_{{NN(1,2);} \atop {(0,11)}}$ & 0  \\
 
 0 & $\bar{Z}^{23(1,1)}_{{NN(1,1);} \atop {(1,00)}}$ & 0 & $\bar{Z}^{23(1,1)}_{{NN(1,2);} \atop {(1,00)}}$   \\

$\bar{Z}^{23(1,1)}_{{NN(2,1);} \atop {(0,11)}}$ & 0 & $\bar{Z}^{23(1,1)}_{{NN(2,2);} \atop {(0,11)}}$ & 0   \\

 0 & $\bar{Z}^{23(1,1)}_{{NN(2,1);} \atop {(1,00)}}$ & 0 & $\bar{Z}^{23(1,1)}_{{NN(2,2);} \atop {(1,00)}}$    \\
\end{tabular}
\right)
\end{equation}
%---------------------------------------------------------------------------------------------------------------------------------------------------------------
\begin{equation}
\label{Z31NN}
\bar{Z}^{31}_{NN} = \left(
\begin{tabular}{cccc}
$\bar{Z}^{31(1,1)}_{{NN(1,1);} \atop {(0,11)}}$ & 0 & $\bar{Z}^{31(1,1)}_{{NN(1,2);} \atop {(0,11)}}$ & 0 \\

 0 & $\bar{Z}^{31(1,1)}_{{NN(1,1);} \atop {(1,00)}}$ & 0 & $\bar{Z}^{31(1,1)}_{{NN(1,2);} \atop {(1,00)}}$ \\

$\bar{Z}^{31(1,1)}_{{NN(2,1);} \atop {(0,11)}}$ & 0 & $\bar{Z}^{31(1,1)}_{{NN(2,2);} \atop {(0,11)}}$ & 0  \\

0 & $\bar{Z}^{31(1,1)}_{{NN(2,1);} \atop {(1,00)}}$ & 0 & $\bar{Z}^{31(1,1)}_{{NN(2,2);} \atop {(1,00)}}$ 
\end{tabular}
\right)
\end{equation}
%---------------------------------------------------------------------------------------------------------------------------------------------------------------
\begin{equation}
\label{Z32NN}
\bar{Z}^{32}_{NN} = \left(
\begin{tabular}{cccc}
 $\bar{Z}^{32(1,1)}_{{NN(1,1);} \atop {(0,11)}}$ & 0 & $\bar{Z}^{32(1,1)}_{{NN(1,2);} \atop {(0,11)}}$ & 0 \\
 
 0 & $\bar{Z}^{32(1,1)}_{{NN(1,1);} \atop {(1,00)}}$ & 0 & $\bar{Z}^{32(1,1)}_{{NN(1,2);} \atop {(1,00)}}$  \\

 $\bar{Z}^{32(1,1)}_{{NN(2,1);} \atop {(0,11)}}$ & 0 & $\bar{Z}^{32(1,1)}_{{NN(2,2);} \atop {(0,11)}}$ & 0   \\

 0 & $\bar{Z}^{32(1,1)}_{{NN(2,1);} \atop {(1,00)}}$ & 0 & $\bar{Z}^{32(1,1)}_{{NN(2,2);} \atop {(1,00)}}$ 
\end{tabular}
\right)
\end{equation}
and three elements with $\alpha = \bar{K}N$:
%---------------------------------------------------------------------------------------------------------------------------------------------------------------
\begin{equation}
\label{Z22KN}
\bar{Z}^{22}_{\bar{K}N} = \left(
\begin{tabular}{cccc}
$\bar{Z}^{22(1,1)}_{{\bar{K}N(1,1);} \atop {(0,00)}}$ & 0 & $\bar{Z}^{22(1,1)}_{{\bar{K}N(1,1);} \atop {(0,01)}}$ & 0 \\
 0 &
$\bar{Z}^{22(1,1)}_{{\bar{K}N(1,1);} \atop {(1,00)}}$ & 0 & $\bar{Z}^{22(1,1)}_{{\bar{K}N(1,1);} \atop {(1,01)}}$   \\
$\bar{Z}^{22(1,1)}_{{\bar{K}N(1,1);} \atop {(0,10)}}$ & 0 & $\bar{Z}^{22(1,1)}_{{\bar{K}N(1,1);} \atop {(0,11)}}$ & 0   \\
 0 &
$\bar{Z}^{22(1,1)}_{{\bar{K}N(1,1);} \atop {(1,10)}}$ & 0 & $\bar{Z}^{22(1,1)}_{{\bar{K}N(1,1);} \atop {(1,11)}}$ 
\end{tabular}
\right)
\end{equation}
%---------------------------------------------------------------------------------------------------------------------------------------------------------------
\begin{equation}
\label{Z23KN}
\bar{Z}^{23}_{\bar{K}N} = \left(
\begin{tabular}{cc}
$\bar{Z}^{23(1,1)}_{{\bar{K}N(1,1);} \atop {(0,01)}}$ & 0  \\
 0 & $\bar{Z}^{23(1,1)}_{{\bar{K}N(1,1);} \atop {(1,00)}}$  \\
 $\bar{Z}^{23(1,1)}_{{\bar{K}N(1,1);} \atop {(0,11)}}$ & 0  \\
 0 & $\bar{Z}^{23(1,1)}_{{\bar{K}N(1,1);} \atop {(1,10)}}$ 
\end{tabular}
\right)
\end{equation}
%---------------------------------------------------------------------------------------------------------------------------------------------------------------
\begin{equation}
\label{Z32KN}
\bar{Z}^{32}_{\bar{K}N} = \left(
\begin{tabular}{cccc}
$\bar{Z}^{32(1,1)}_{{\bar{K}N(1,1);} \atop {(0,10)}}$ & 0 & $\bar{Z}^{32(1,1)}_{{\bar{K}N(1,1);} \atop {(0,11)}}$ & 0 \\
 0 &
$\bar{Z}^{32(1,1)}_{{\bar{K}N(1,1);} \atop {(1,00)}}$ & 0 & $\bar{Z}^{32(1,1)}_{{\bar{K}N(1,1);} \atop {(1,01)}}$  
\end{tabular}
\right)
\end{equation}

Elements of $\bar{\tau}^{\rho}_{\alpha \beta}$ matrix Eq.(\ref{tau5x5}) are parts of the
$NNN$ subsystem ($\rho = 1$):
%---------------------------------------------------------------------------------------------------------------------------------------------------------------
\begin{equation}
\label{tau1NNNN}
\bar{\tau}^{1}_{NN,NN} = \left(
\begin{tabular}{cccc}
 $\bar{\tau}^{1(1)}_{{NN,NN(1,1);} \atop {(00,11)}}$ & $\bar{\tau}^{1(1)}_{{NN,NN(1,1);} \atop {(01,10)}}$ &
  $\bar{\tau}^{1(1)}_{{NN,NN(1,2);} \atop {(00,11)}}$  & $\bar{\tau}^{1(1)}_{{NN,NN(1,2);} \atop {(01,10)}}$  \\
 
 $\bar{\tau}^{1(1)}_{{NN,NN(1,1);} \atop {(10,01)}}$  & $\bar{\tau}^{1(1)}_{{NN,NN(1,1);} \atop {(11,00)}}$  & 
  $\bar{\tau}^{1(1)}_{{NN,NN(1,2);} \atop {(10,01)}}$  & $\bar{\tau}^{1(1)}_{{NN,NN(1,2);} \atop {(11,00)}}$   \\

 $\bar{\tau}^{1(1)}_{{NN,NN(2,1);} \atop {(00,11)}}$ & $\bar{\tau}^{1(1)}_{{NN,NN(2,1);} \atop {(01,10)}}$ &
  $\bar{\tau}^{1(1)}_{{NN,NN(2,2);} \atop {(00,11)}}$  & $\bar{\tau}^{1(1)}_{{NN,NN(2,2);} \atop {(01,10)}}$  \\
 
 $\bar{\tau}^{1(1)}_{{NN,NN(2,1);} \atop {(10,01)}}$  & $\bar{\tau}^{1(1)}_{{NN,NN(2,1);} \atop {(11,00)}}$  & 
  $\bar{\tau}^{1(1)}_{{NN,NN(2,2);} \atop {(10,01)}}$  & $\bar{\tau}^{1(1)}_{{NN,NN(2,2);} \atop {(11,00)}}$  
\end{tabular}
\right)
\end{equation}
of the $\bar{K}NN$ subsystem ($\rho = 2$):
%---------------------------------------------------------------------------------------------------------------------------------------------------------------
\begin{equation}
\label{tau2NNNN}
\bar{\tau}^{2}_{NN,NN} = \left(
\begin{tabular}{cccc}
 $\bar{\tau}^{2(1)}_{{NN,NN(1,1);} \atop {(00,11)}}$ & 0 &
   $\bar{\tau}^{2(1)}_{{NN,NN(1,2);} \atop {(00,11)}}$  & 0  \\
 
 0  & $\bar{\tau}^{2(1)}_{{NN,NN(1,1);} \atop {(11,00)}}$  & 
  0  & $\bar{\tau}^{2(1)}_{{NN,NN(1,2);} \atop {(11,00)}}$   \\

 $\bar{\tau}^{2(1)}_{{NN,NN(2,1);} \atop {(00,11)}}$ & 0 &
  $\bar{\tau}^{2(1)}_{{NN,NN(2,2);} \atop {(00,11)}}$  & 0  \\
 
 0 & $\bar{\tau}^{2(1)}_{{NN,NN(2,1);} \atop {(11,00)}}$  & 
  0  & $\bar{\tau}^{2(1)}_{{NN,NN(2,2);} \atop {(11,00)}}$  
\end{tabular}
\right)
\end{equation}
%---------------------------------------------------------------------------------------------------------------------------------------------------------------
\begin{equation}
\label{tau2NNKN}
\bar{\tau}^{2}_{NN,\bar{K}N} = \left(
\begin{tabular}{cccc}
  0 & 0 &
  $\bar{\tau}^{2(1)}_{{NN,\bar{K}N(1,1);} \atop {(00,11)}}$  & $\bar{\tau}^{2(1)}_{{NN,\bar{K}N(1,1);} \atop {(01,11)}}$  \\
 
 $\bar{\tau}^{2(1)}_{{NN,\bar{K}N(1,1);} \atop {(10,00)}}$  & $\bar{\tau}^{2(1)}_{{NN,\bar{K}N(1,1);} \atop {(11,00)}}$  & 
  0 & 0   \\

 0 & 0 &
  $\bar{\tau}^{2(1)}_{{NN,\bar{K}N(2,1);} \atop {(00,11)}}$  & $\bar{\tau}^{2(1)}_{{NN,\bar{K}N(2,1);} \atop {(01,11)}}$  \\
 
 $\bar{\tau}^{2(1)}_{{NN,\bar{K}N(2,1);} \atop {(10,00)}}$  & $\bar{\tau}^{2(1)}_{{NN,\bar{K}N(2,1);} \atop {(11,00)}}$  & 
 0 & 0 
\end{tabular}
\right)
\end{equation}
%---------------------------------------------------------------------------------------------------------------------------------------------------------------
\begin{equation}
\label{tau2KNNN}
\bar{\tau}^{2}_{\bar{K}N,NN} = \left(
\begin{tabular}{cccc}
 0 & $\bar{\tau}^{2(1)}_{{\bar{K}N,NN(1,1);} \atop {(01,00)}}$ & 
  0  & $\bar{\tau}^{2(1)}_{{\bar{K}N,NN(1,2);} \atop {(01,00)}}$  \\
 
 0 & $\bar{\tau}^{2(1)}_{{\bar{K}N,NN(1,1);} \atop {(11,00)}}$  & 
  0 & $\bar{\tau}^{2(1)}_{{\bar{K}N,NN(1,2);} \atop {(11,00)}}$   \\

 $\bar{\tau}^{2(1)}_{{\bar{K}N,NN(1,1);} \atop {(00,11)}}$ & 0 &
  $\bar{\tau}^{2(1)}_{{\bar{K}N,NN(1,2);} \atop {(00,11)}}$  & 0  \\
 
 $\bar{\tau}^{2(1)}_{{\bar{K}N,NN(1,1);} \atop {(10,11)}}$  & 0  & 
  $\bar{\tau}^{2(1)}_{{\bar{K}N,NN(1,2);} \atop {(10,11)}}$  & 0
\end{tabular}
\right)
\end{equation}
%---------------------------------------------------------------------------------------------------------------------------------------------------------------
\begin{equation}
\label{tau2KNKN}
\bar{\tau}^{2}_{\bar{K}N,\bar{K}N} = \left(
\begin{tabular}{cccc}
 $\bar{\tau}^{2(1)}_{{\bar{K}N,\bar{K}N(1,1);} \atop {(00,00)}}$ & $\bar{\tau}^{2(1)}_{{\bar{K}N,\bar{K}N(1,1);} \atop {(01,00)}}$ &
  0 & 0 \\
 
 $\bar{\tau}^{2(1)}_{{\bar{K}N,\bar{K}N(1,1);} \atop {(10,00)}}$  & $\bar{\tau}^{2(1)}_{{\bar{K}N,\bar{K}N(1,1);} \atop {(11,00)}}$  & 
  0 & 0   \\

 0 & 0 &
  $\bar{\tau}^{2(1)}_{{\bar{K}N,\bar{K}N(1,1);} \atop {(00,11)}}$  & $\bar{\tau}^{2(1)}_{{\bar{K}N,\bar{K}N(1,1);} \atop {(01,11)}}$  \\
 
 0 & 0  & 
  $\bar{\tau}^{2(1)}_{{\bar{K}N,\bar{K}N(1,1);} \atop {(10,11)}}$  & $\bar{\tau}^{2(1)}_{{\bar{K}N,\bar{K}N(1,1);} \atop {(11,11)}}$  
\end{tabular}
\right)
\end{equation}
and of the $\bar{K}N + NN$ partition ($\rho = 3$):
%---------------------------------------------------------------------------------------------------------------------------------------------------------------
\begin{equation}
\label{tau3NNNN}
\bar{\tau}^{3}_{NN,NN} = \left(
\begin{tabular}{cccc}
 $\bar{\tau}^{3(1)}_{{NN,NN(1,1);} \atop {(00,11)}}$ & 0 &
  $\bar{\tau}^{3(1)}_{{NN,NN(1,2);} \atop {(00,11)}}$  & 0  \\
 
 0 & $\bar{\tau}^{3(1)}_{{NN,NN(1,1);} \atop {(11,00)}}$  & 
  0  & $\bar{\tau}^{3(1)}_{{NN,NN(1,2);} \atop {(11,00)}}$   \\

 $\bar{\tau}^{3(1)}_{{NN,NN(2,1);} \atop {(00,11)}}$ & 0 &
  $\bar{\tau}^{3(1)}_{{NN,NN(2,2);} \atop {(00,11)}}$  & 0  \\
 
 0  & $\bar{\tau}^{3(1)}_{{NN,NN(2,1);} \atop {(11,00)}}$  & 
  0  & $\bar{\tau}^{3(1)}_{{NN,NN(2,2);} \atop {(11,00)}}$  
\end{tabular}
\right)
\end{equation}
%---------------------------------------------------------------------------------------------------------------------------------------------------------------
\begin{equation}
\label{tau3NNKN}
\bar{\tau}^{3}_{NN,\bar{K}N} = \left(
\begin{tabular}{cc}
 $\bar{\tau}^{3(1)}_{{NN,\bar{K}N(1,1);} \atop {(00,11)}}$ & 0  \\
 
 0 & $\bar{\tau}^{3(1)}_{{NN,\bar{K}N(1,1);} \atop {(11,00)}}$   \\

 $\bar{\tau}^{3(1)}_{{NN,\bar{K}N(2,1);} \atop {(00,11)}}$ & 0  \\
 
 0  & $\bar{\tau}^{3(1)}_{{NN,\bar{K}N(2,1);} \atop {(11,00)}}$    
\end{tabular}
\right)
\end{equation}
%---------------------------------------------------------------------------------------------------------------------------------------------------------------
\begin{equation}
\label{tau3KNNN}
\bar{\tau}^{3}_{\bar{K}N,NN} = \left(
\begin{tabular}{cccc}
 $\bar{\tau}^{3(1)}_{{\bar{K}N,NN(1,1);} \atop {(00,11)}}$ & 0 &
  $\bar{\tau}^{3(1)}_{{\bar{K}N,NN(1,2);} \atop {(00,11)}}$  & 0  \\
 
 0 & $\bar{\tau}^{3(1)}_{{\bar{K}N,NN(1,1);} \atop {(11,00)}}$  & 
  0 & $\bar{\tau}^{3(1)}_{{\bar{K}N,NN(1,2);} \atop {(11,00)}}$   
\end{tabular}
\right)
\end{equation}
%---------------------------------------------------------------------------------------------------------------------------------------------------------------
\begin{equation}
\label{tau3KNKN}
\bar{\tau}^{3}_{\bar{K}N,\bar{K}N} = \left(
\begin{tabular}{cc}
 $\bar{\tau}^{3(1)}_{{\bar{K}N,\bar{K}N(1,1);} \atop {(00,11)}}$ & 0 \\
 
 0  & $\bar{\tau}^{3(1)}_{{\bar{K}N,\bar{K}N(1,1);} \atop {(11,00)}}$  
\end{tabular}
\right)
\end{equation}

%\begin{thebibliography}{00}  %for 2 digits

\end{document}